\documentclass[aps,prl,reprint,superscriptaddress,floatfix,preprintnumbers]{revtex4-2}

\pdfoutput=1

\usepackage[T1]{fontenc}
\usepackage[utf8]{inputenc}
\usepackage{amsmath,amssymb,amsfonts,mathtools,bm}
\usepackage[colorlinks=true, citecolor=blue, linkcolor=blue]{hyperref}
\usepackage[final]{graphicx}
\usepackage{xcolor}
\usepackage{booktabs}
\usepackage{relsize}
\usepackage{lineno}
\usepackage{comment}
\usepackage{braket}
\usepackage{dsfont}
\usepackage{float}
\usepackage{hyperref}
\usepackage{needspace}
\usepackage{enumitem}
\usepackage{braket}

\usepackage{lineno}

\usepackage{csquotes}

\usepackage{algorithm}
\usepackage{algpseudocode}

\def\Babar{{\mbox{\slshape B\kern-0.1em{\smaller A}\kern-0.1em B\kern-0.1em{\smaller A\kern-0.2em R}}}}

\usepackage{cancel}

\usepackage{soul}

\usepackage{siunitx} 
\usepackage[caption=false]{subfig}

\setstcolor{red}

\begin{document}

\preprint{TUM-EFT 219/26}

\title{How Quantum Is Bottomonium in the Quark-Gluon Plasma?}

\author{Nora Brambilla}
\email{nora.brambilla@tum.de}
\affiliation{Technical University of Munich, TUM School of Natural Sciences, Physics Department, James-Franck-Strasse 1, 85748 Garching, Germany}
\affiliation{Institute for Advanced Study, Technische Universität München, Lichtenbergstrasse 2 a, 85748 Garching, Germany}
\affiliation{Munich Data Science Institute, Technische Universität München, Walther-von-Dyck-Strasse 10, 85748 Garching, Germany}

\author{Tom Magorsch}
\email{tom.magorsch@tum.de}
\affiliation{Technical University of Munich, TUM School of Natural Sciences, Physics Department, James-Franck-Strasse 1, 85748 Garching, Germany}

\begin{abstract}
We study how much quantum structure bottomonium retains during its evolution in the quark-gluon plasma. Within the open quantum system framework, we simulate the Lindblad equation obtained from potential nonrelativistic QCD in the quantum Brownian regime. From the resulting real-time evolution of the bottomonium density matrix, we compute the Wigner transform and study its negativity as a measure of nonclassicality and a test of a key condition underlying classical Langevin-type approximations. We find that the medium suppresses the negativity, which nevertheless approaches a finite plateau close to that of the $1\mathrm{S}$ state. The overlaps with the excited $2\mathrm{S}$ and $3\mathrm{S}$ states arise from near-canceling positive and negative phase-space contributions, driven by negative regions in the Wigner transforms of the bound-state projectors, while the $1\mathrm{S}$ overlap is insensitive to such contributions. We further show that the position-space coherence is similarly suppressed and plateaus close to the 1S value. We trace the finite residual quantum structure to conditional
purification: the medium preferentially dissolves weakly bound and unbound modes with large $\braket{r^2}$, driving the surviving singlet $\mathrm{S}$-wave ensemble toward the $1\mathrm{S}$ state. More broadly, these results reveal that classicalization is partial and observable dependent, a feature that is relevant across applications of open quantum system methods in fundamental physics.
\end{abstract}

\maketitle

\textbf{\textit{Introduction.---}} 
The evolution of bound states in the quark-gluon plasma (QGP) is often described by classical transport equations like rate, Boltzmann, or Langevin-type equations~\cite{Du:2015wha,Yao:2020xzw,Young:2008he,Oei:2024pva,Andronic:2024oxz}. For heavy quarkonium, in recent years, the combination of non-relativistic effective field theories with the open quantum system framework enabled a full quantum mechanical description of the non-equilibrium evolution from first principles~\cite{Akamatsu:2014qsa,Blaizot:2015hya,Katz:2015qja,Brambilla:2016wgg,Brambilla:2017zei,Delorme:2024rdo,Akamatsu:2020ypb,Yao:2021lus,Rothkopf:2019ipj}. The presence of these two approaches raises the question whether a quantum mechanical treatment of quarkonium is necessary, or if a classical transport equation can be justified. In the open quantum system framework, controlled classical approximations to quantum master equations exist. In particular, in the quantum Brownian regime, it is possible to obtain a Langevin equation, while the quantum optical regime gives rise to a Boltzmann equation~\cite{Calzetta:2000tp,Blaizot:2017ypk,spohn,Yao:2018nmy,theoryofoqs}. To obtain a Langevin equation from the corresponding Lindblad equation in the quantum Brownian regime, one performs a Wigner transformation and treats the resulting distribution as a classical probability distribution, assuming it is nonnegative. The question of when a classical description is adequate arises broadly in the application of open quantum system methods across fundamental physics, which describe decoherence and dissipation in settings from heavy-ion collisions~\cite{Yao:2021lus,Akamatsu:2020ypb} and neutrino oscillations~\cite{Benatti:2000ph,Nieves:2020jjg,DeRomeri:2023dht} to the evolution of quantum fields and gravity in the early universe~\cite{Martin:2018zbe,Colas:2022hlq,Burgess:2022nwu,Salcedo:2024smn,Salcedo:2025ezu,Ireland:2026txt}.

In this letter, we study bottomonium in a strongly coupled medium utilizing potential non-relativistic QCD (pNRQCD). Within this framework, a Lindblad equation has been derived in the quantum Brownian regime, enabling phenomenological predictions for bottomonium suppression in heavy-ion collision experiments~\cite{Brambilla:2016wgg,Brambilla:2022ynh,Brambilla:2023hkw,Brambilla:2024tqg}. While it is expected that the interaction with the medium leads to decoherence and some form of classicalization, the question of whether the evolution can be accurately described by a corresponding classical transport remains open. 
In a one-dimensional abelian model, the Lindblad evolution has recently been compared to its semiclassical Fokker-Planck equation, which was found to reproduce the quantum yields accurately, with the Wigner distribution of the full density matrix becoming close to positive~\cite{Daddi-Hammou:2025hdz}. Here we work in three dimensions, with the Coulombic pNRQCD potentials and the full color structure, and characterize the surviving color-singlet, low-angular-momentum ensemble that carries the $\mathrm{S}$-wave yields, rather than the full density matrix. We thus set out to answer the question: \textit{How much quantum structure does bottomonium retain throughout its evolution within the QGP, and what impact does it have on relevant observables?}

We find that the medium suppresses both Wigner negativity and position-space coherence, but leaves a finite residual quantum structure. Its phenomenological relevance is observable-dependent: while the $1\mathrm{S}$ survival probability is largely insensitive, the $2\mathrm{S}$ and $3\mathrm{S}$ projections remain sensitive to negative regions of the bound-state Wigner distributions.
The conditional state becomes increasingly $1\mathrm{S}$-dominated through conditional purification, as the medium predominantly dissolves weakly bound and unbound modes, causing residual negativity and coherence to approach $1\mathrm{S}$ values.

\textbf{\textit{Quarkonium in medium as open quantum system.---}}
Exploiting the heavy quark mass, heavy quarkonium can be described non-relativistically. In particular, bottomonium exhibits a strong separation of scales, as the radius of the bound state is small, making the application of effective field theories (EFTs) particularly suitable. Utilizing this hierarchy of scales, Refs.~\cite{Brambilla:2016wgg,Brambilla:2017zei} used pNRQCD~\cite{Brambilla:1999xf} to derive a quantum master equation for the real-time evolution of the quarkonium density matrix in a color medium. Further assuming that the binding energy $E$ is smaller than the relevant thermal scale $\pi T$, the master equation can be brought into Lindblad form~\cite{Brambilla:2022ynh,Lindblad:1975ef,Gorini:1975nb}.
This hierarchy corresponds to the quantum Brownian regime. 
At next-to-leading order in an expansion in $E/(\pi T)$, the evolution equation can thus be written as~\cite{Brambilla:2022ynh}
\begin{equation}
    \frac{\mathrm{d}\rho}{\mathrm{d}t} = -\mathrm{i}[H,\rho] + \sum^5_{n=0}\left(L_n\rho L^\dagger_n-\frac{1}{2}\left\{L^\dagger_nL_n,\rho\right\}\right).
    \label{eq:lind}
\end{equation}
Here $\rho$ is the density matrix of the bottomonium, $H$ the Hamiltonian of the evolution, and the six jump operators $L_n$ contain the effects of the medium on the bound state. We provide more details on the Lindblad equation in the Supplemental Material. Assuming isotropy, the density matrix $\rho$ is block diagonal in color and angular momentum
\begin{equation}
\rho \;=\;
\begin{pmatrix}
\rho_s & 0\\
0 & \rho_o
\end{pmatrix},
\qquad
\rho_{c} \;=\; \bigoplus_{\ell}\rho_{c}^{(\ell)},
\end{equation}  
with $\rho^{(\ell)}_{c}$ a block with angular momentum $\ell$ and color $c$. The jump operators $L_n$ induce transitions between different color and angular momentum sectors governed by the pNRQCD dipole interaction. Consequently, the trace of a block $\rho^{(\ell)}_c$ is not conserved as the medium-induced transitions can deplete the sector. To separate this depletion of the ensemble from the quantum structure of the surviving component, we define the \textit{conditional density matrix} $\rho_{c|\ell_\text{max}}$ as 
\begin{equation}
    \rho_{c|\ell_\text{max}} = \frac{1}{Z_{c|\ell_\text{max}}}\bigoplus^{\ell_\text{max}}_{\ell=0}\rho^{(\ell)}_c, \qquad Z_{c|\ell_\text{max}}=\sum^{\ell_\text{max}}_{\ell=0}\text{Tr}\rho^{(\ell)}_c.
\end{equation}

Using the open-source code QTraj~\cite{Omar:2021kra}, we simulate the evolution of bottomonium in a Bjorken medium~\cite{PhysRevD.27.140} from a temperature of  $T_0=425\,$MeV at $\tau_0=0.6\,$fm down to $190\,$MeV. We use the leading order Coulombic singlet and octet potentials with $M=4.73\,\mathrm{GeV}$ and $\alpha_s=0.468$ taken from previous studies~\cite{Brambilla:2022ynh}, and parametrize the transport coefficients as $\kappa=\hat\kappa T^3$ and $\gamma=\hat\gamma T^3$ with $\hat\kappa=4$ and $\hat\gamma=0$, consistent with previous phenomenological results~\cite{Brambilla:2023hkw}.
We simulate the Lindblad evolution with the Monte Carlo wave-function method~\cite{PhysRevLett.68.580} implemented in QTraj~\cite{Omar:2021kra}.
We initialize the wave function $u(r)$ as a color singlet compact Gaussian $u(r)\propto r\exp(-r^2/(0.2a_0)^2)$, with $a_0$ the Bohr radius, and evolve it with the vacuum Hamiltonian until $\tau_0$, where we turn on the full Lindblad evolution. By averaging over the wave functions of about $\num{2e5}$ trajectories, we reconstruct the full real-time evolution of the density matrix $\rho(t)$. More details on the computational setup are provided in Ref.~\cite{Brambilla:2022ynh} and the Supplemental Material.

\begin{figure}
    \centering
    \includegraphics[width=0.95\linewidth]{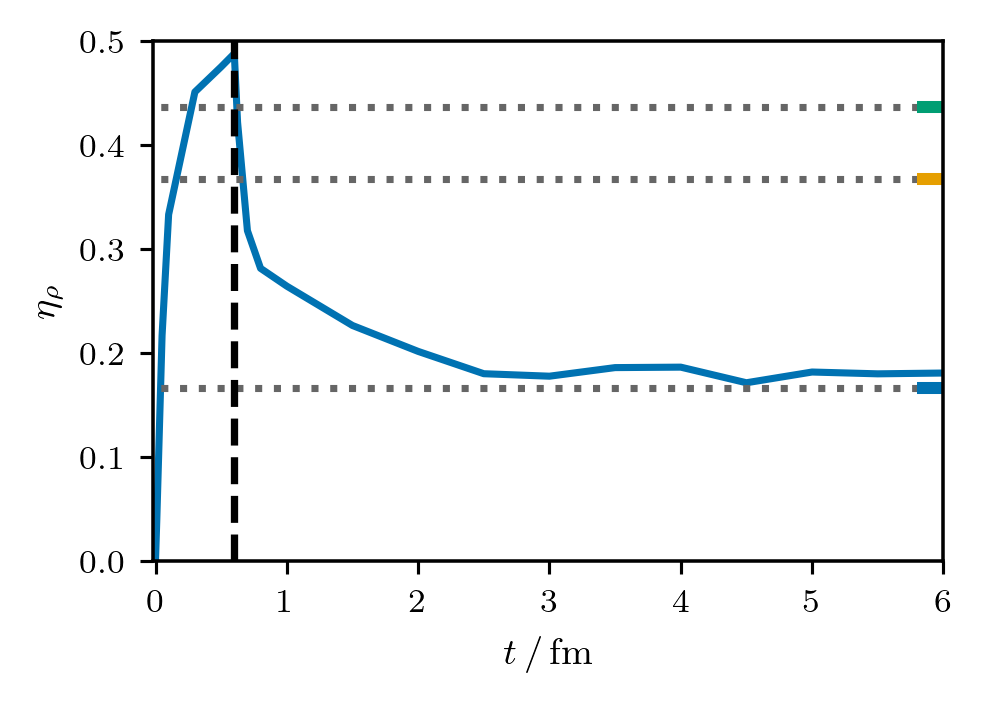}
    \caption{The time evolution of the negativity $\eta_\rho$ for the conditional $\ell=0$ singlet state $\rho_{s|0}$ within a Bjorken medium. The state is initialized as a compact Gaussian and evolved with the vacuum Hamiltonian until $0.6\,$fm, where the thermal medium is switched on as indicated by the black dashed line. The dotted horizontal lines indicate the negativity of the $1\mathrm{S}$ (blue), $2\mathrm{S}$ (orange) and $3\mathrm{S}$ (green) bound states.}
    \label{fig:1_negativity}
\end{figure}

\textbf{\textit{Partial classicalization of bottomonium in the QGP.---}}
To obtain a Langevin equation from the Lindblad equation, in the quantum Brownian regime, one performs a Wigner transformation of the master equation. The Wigner transform~\cite{Wigner:1932eb} of the density matrix $\rho_c(\vec{r},\vec{r}^\prime)$ with color $c$ is given by
\begin{equation}
W_{\rho_c}(\vec{R},\vec{P})=\frac{1}{(2\pi)^3}\int \mathrm{d}^3s\, \mathrm{e}^{-\mathrm{i}\vec{P}\cdot\vec{s}}\rho_c\left(\vec{R}-\frac{\vec{s}}{2},\vec{R}+\frac{\vec{s}}{2}\right).
\label{eq:wig}
\end{equation}
As we assume isotropy, $W_{\rho_c}$ generally only depends on $|\vec{R}|$, $|\vec{P}|$ and $\theta_{RP}=\angle(\vec{R},\vec{P})$. We provide details on the evaluation of the Wigner transformation in the Supplemental Material. After the Wigner transformation, one identifies the resulting distribution $W(\vec{R},\vec{P})$ with a classical probability distribution. Truncating the Moyal expansion~\cite{Moyal_1949,Blaizot:2017ypk}, this classical distribution obeys a Fokker-Planck equation, which in turn can be written as a Langevin equation.
To make this identification, it is necessary to assume that the Wigner distribution $W(\vec{R},\vec{P})$ is nonnegative.
Since our simulations provide access to the full density matrix $\rho(t)$, we can directly assess how well this assumption holds.
To this end, we define the negativity $\eta$ as a fraction of negative parts of the Wigner distribution
\begin{equation}
    \eta = \frac{\int_{W<0} \mathrm{d}^3R\,\mathrm{d}^3P\;\left|W\left(\vec{R},\vec{P}\right)\right|}{\int\mathrm{d}^3R\,\mathrm{d}^3P\;\left|W\left(\vec{R},\vec{P}\right)\right|}.
\end{equation}
The negativity is an established measure of nonclassicality. We note that our definition is related to the more common Kenfack–Życzkowski nonclassical volume~\cite{Kenfack:2004ges} by a monotonic transformation.
In quantum information, Wigner negativity is treated as a resource, as nonnegative phase-space representations of the state, the operations applied to it, and the measurements used to probe it are sufficient for an efficient classical simulation~\cite{Veitch:2012ttw,Mari:2012ypq,Rahimi-Keshari:2016nas}. Classicality in this sense is therefore a property of the state, the operations, and the measurements taken together, rather than of the state alone. For pure states, Hudson's theorem implies that a state has zero negativity if and only if it is Gaussian~\cite{HUDSON1974249,Soto:1983avg}.
From the real-time evolution of the density matrix, we obtain the Wigner distribution of the conditional $\ell=0$ singlet state $\rho_{s|0}$, which carries the S-wave position-space overlaps, by numerically integrating Eq.~\eqref{eq:wig}.
In Fig.~\ref{fig:1_negativity} we show the resulting time evolution of the negativity $\eta_\rho$ for the $\ell=0$ color singlet sector.
By Hudson’s theorem, the initial pure Gaussian state has zero negativity, while the non-Gaussianity generated by the Coulombic vacuum evolution leads to the observed steep rise in negativity as the state remains pure.
When the medium is turned on at $0.6\,$fm, it quickly suppresses the negativity, leading to nearly exponential decay. At about $3\,$fm the negativity plateaus to approximately $18\%$. We compare $\eta_\rho$ to the negativity of the bound states themselves, by calculating the Wigner transformation $W_{n\mathrm{S}}$ of the bound state projectors $\Pi_{n\mathrm{S}}=\ket{n\mathrm{S}}\bra{n\mathrm{S}}$. Here, the $\ket{nS}$ are the lowest S-wave eigenstates of the non-relativistic
vacuum Hamiltonian. We find the corresponding negativities $\eta_{1S}=16.6\%$, $\eta_{2S}=36.7\%$, $\eta_{3S}=43.6\%$, shown as dotted lines in Fig.~\ref{fig:1_negativity}. 
In Ref.~\cite{Daddi-Hammou:2025hdz} the negativity of the full density matrix was found to become small at late times in a one-dimensional abelian model. The comparison is not direct, as we work in three dimensions and evaluate $\eta$ for the conditional $\ell = 0$ singlet density matrix relevant for the S-wave overlaps, and the dimensionality, the color structure, and the potential further affect the phase-space structure of the surviving state.

We expect neglected pre-equilibrium effects before $0.6\,$fm to mainly affect the transient maximum. The coherent evolution of the initial Gaussian state will generate negativity despite possible pre-equilibrium effects, while the late-time plateau is increasingly governed by
medium-induced filtering toward the compact low-lying singlet component and should therefore be largely insensitive to the details of the early-time evolution.

\begin{figure}
    \centering
    \includegraphics[width=0.95\linewidth]{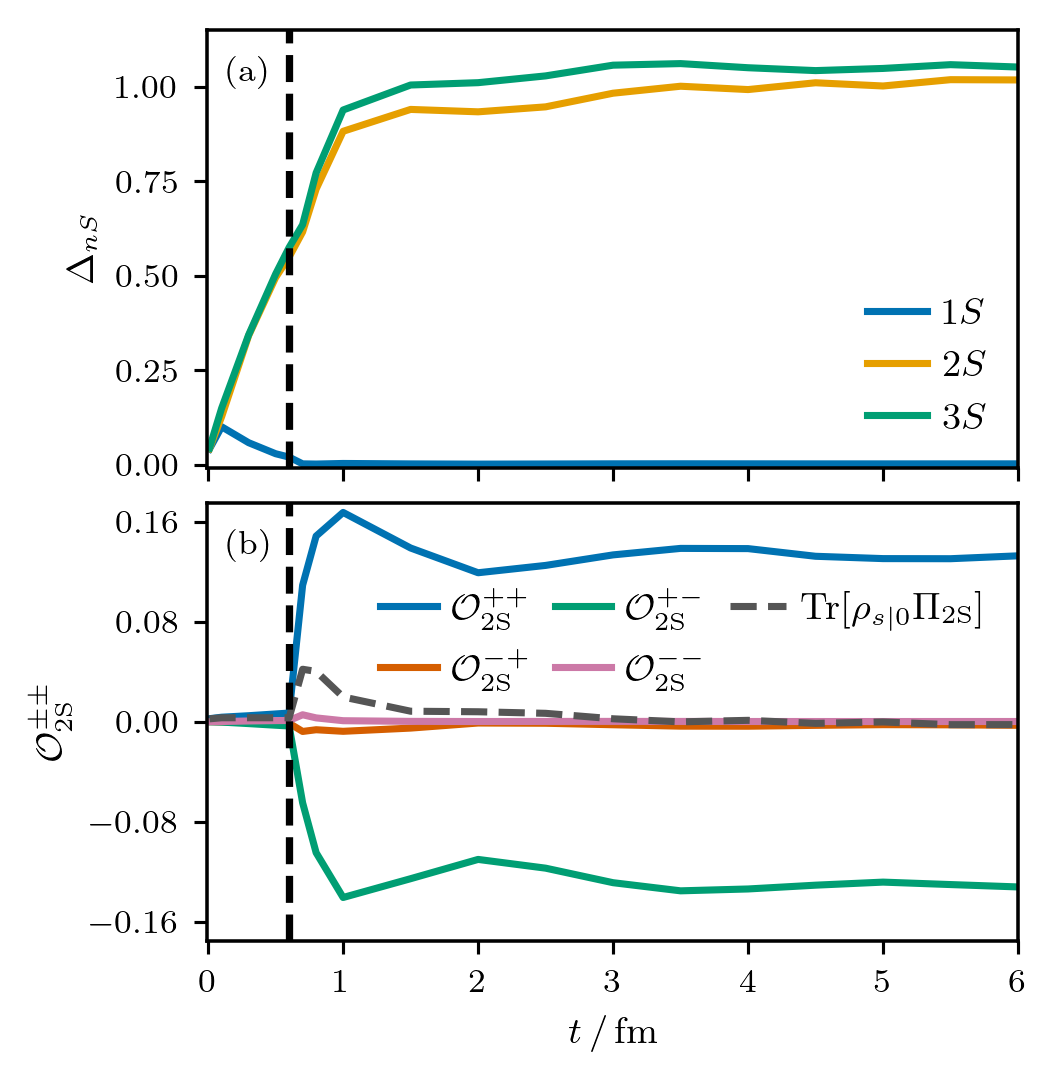}
    \caption{\textit{(a)} The time evolution of the sensitivity $\Delta_{n\mathrm{S}}$ of the bound-state overlaps $\text{Tr}[\rho_{s|0}\Pi_{n\mathrm{S}}]$ to the negative regions of the Wigner distribution. \textit{(b)} The time evolution of the individual terms in the overlap decomposition $\mathcal{O}^{\pm\pm}_{2\mathrm{S}}$ for the $2\mathrm{S}$ state including their sum $\mathrm{Tr}[\rho_{s|0}\Pi_{2\mathrm{S}}]$.}
    \label{fig:2sensitivity}
\end{figure}

While the sizable negativity questions the approximation of the Wigner function by a classical probability distribution, it is not clear how relevant this is to the observables of interest. For bottomonium, these are the survival probabilities, which are proportional to the overlaps with the vacuum bound-state projectors $\text{Tr}[\rho \Pi_{n\mathrm{S}}]$. Since $\Pi_{n\mathrm{S}}$ projects on the $\ell=0$ color singlet sector, we perform the analysis with the overlaps of the conditional state $\mathrm{Tr}[\rho_{s|0}\Pi_{n\mathrm{S}}]$, which differ only by the common normalization factor $Z_{s|0}$. To assess their sensitivity to negative regions of the Wigner distribution, we write
\begin{align}
    \text{Tr}[\rho_{s|0}\Pi_{n\mathrm{S}}]=(2\pi)^3\int \mathrm{d}^3R\,\mathrm{d}^3P \; W_{\rho_{s|0}}\left(\vec{R},\vec{P}\right)W_{n\mathrm{S}}\left(\vec{R},\vec{P}\right),
\end{align}
where $W_{\rho_{s|0}}$ and $W_{n\mathrm{S}}$ are the Wigner transformations of the density matrix $\rho_{s|0}$ and the projectors $\Pi_{n\mathrm{S}}$ respectively.
Since both $W_{\rho_{s|0}}$ and $W_{n\mathrm{S}}$ have negative regions, the overlaps can be written as a sum of four distinct contributions $\text{Tr}[\rho_{s|0}\Pi_{n\mathrm{S}}]=\mathcal{O}^{++}+\mathcal{O}^{+-}+\mathcal{O}^{-+}+\mathcal{O}^{--}$
with
\begin{equation}
    \mathcal{O}^{\pm\pm} = (2\pi)^3\int\,\mathrm{d}^3R\,\mathrm{d}^3P\; W^{\pm}_{\rho_{s|0}}\left(\vec{R},\vec{P}\right)W^{\pm}_{n\mathrm{S}}\left(\vec{R},\vec{P}\right),
\end{equation}
where $W^{\pm}=(W\pm|W|)/2$ denote the positive and negative parts of $W$.
Here, $\mathcal{O}^{++}$ contributes to the overlaps that arise only from the positive parts of the Wigner distributions.
$\mathcal{O}^{+-}$ and $\mathcal{O}^{-+}$ correspond to the contributions from positive parts of the Wigner distribution of
$\rho_{s|0}$ overlapping with the negative parts of the Wigner distribution of the projector and vice versa.
Lastly, $\mathcal{O}^{--}$ arises from overlaps of the negative contributions to the Wigner distribution of both $\rho_{s|0}$ and the projector $\Pi_{n\mathrm{S}}$. To assess the sensitivity to the negative parts of the Wigner distribution, we define the relative deviation of the overlaps induced by the negative terms as
\begin{equation}
    \Delta_{n\mathrm{S}} = \frac{\left|\mathcal{O}^{++}-\text{Tr}[\rho_{s|0}\Pi_{nS}]\right|}{\mathcal{O}^{++}}.
\end{equation}
We show this sensitivity for the 1S, 2S, and 3S states in panel (a) of Fig.~\ref{fig:2sensitivity}. 
For the $1S$ state $\Delta_{1\mathrm{S}}$ shows a small peak at early times, but quickly decays to practically zero. For the excited states, however, the sensitivity $\Delta_{n\mathrm{S}}$ increases steeply until $1\,$fm, then slowly plateaus. Since, for both excited states, $\Delta_{n\mathrm{S}}$ approaches 1, the excited-state overlaps show significant sensitivity to negative regions of the Wigner distribution. To assess the origin of this sensitivity, we show the individual terms $\mathcal{O}^{\pm\pm}$ of the decomposed overlap integral for the $2S$ projection in panel (b) of Fig.~\ref{fig:2sensitivity}.
We observe that the largest contribution to the overlaps is given by the overlap of the positive regions $\mathcal{O}^{++}_{2\mathrm{S}}$. The overlaps between positive regions in the Wigner transform of $\rho_{s|0}$, with negative parts of the projector $\Pi_{n\mathrm{S}}$ are, however, only slightly smaller in magnitude. By definition $\mathcal{O}^{+-}_{2\mathrm{S}}$ is negative leading to a cancellation between $\mathcal{O}^{++}_{2\mathrm{S}}$ and $\mathcal{O}^{+-}_{2\mathrm{S}}$. Since $\mathcal{O}^{-+}_{2\mathrm{S}}$ and $\mathcal{O}^{--}_{2\mathrm{S}}$ remain small, the excited-state overlaps are small residues of a near-cancellation between $\mathcal{O}^{++}_{2\mathrm{S}}$ and $\mathcal{O}^{+-}_{2\mathrm{S}}$. The negativity relevant for these observables thus mainly resides in the Wigner transform of the measurement projector rather than the evolving state. In the sense of the classicality criterion discussed above, it is thus the nonnegativity requirement on the measurement representation that fails: even for a state with a nonnegative Wigner distribution, $\mathcal{O}^{-+}_{2\mathrm{S}}$ and $\mathcal{O}^{--}_{2\mathrm{S}}$ would vanish while the cancellation between $\mathcal{O}^{++}_{2\mathrm{S}}$ and $\mathcal{O}^{+-}_{2\mathrm{S}}$ would remain. 
Indeed, the semiclassical calculation of Ref.~\cite{Daddi-Hammou:2025hdz} retains this measurement-side quantum structure: the yields are evaluated using the Wigner transforms of the bound-state density matrices, whose $2\mathrm{S}$ Wigner transform contains negative regions and corresponding interference contributions.
Classical phase-space-based transport schemes that identify bound states via positive acceptance criteria lack the $\mathcal{O}^{+-}_{2\mathrm{S}}$ term by construction and thus have no access to the cancellation determining the $2\mathrm{S}$ and $3\mathrm{S}$ yields. Since the physical overlap is the small residue of this cancellation, small inaccuracies in $W_{\rho_{s|0}}$ can translate into $\mathcal{O}(1)$ relative errors in the excited-state yields, whereas no analogous cancellation amplifies inaccuracies in the $1\mathrm{S}$ overlap.

\begin{figure*}[t]
    \centering
    \includegraphics[width=0.95\linewidth]{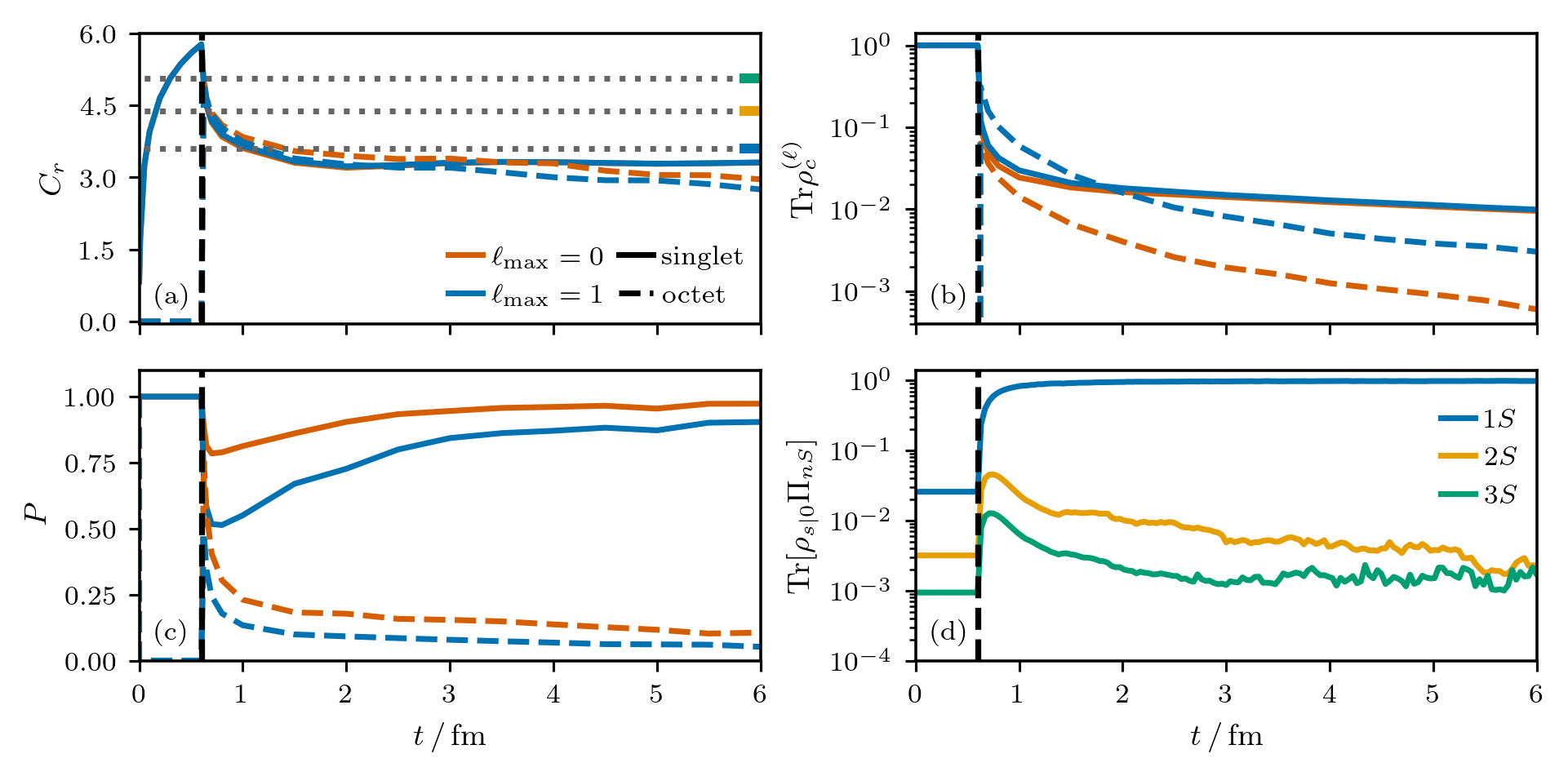}
    \caption{The time evolution of \textit{(a)} the position space coherence $C_r$, (b) the trace of the individual blocks $\mathrm{Tr}\rho^{(\ell)}_c$, (c) the purity $P$ and (d) the overlaps of the conditional color singlet $\ell=0$ sector with the bound states $\mathrm{Tr}[\rho_{s|0}\Pi_{n\mathrm{S}}]$. Panels (a) and (c) involve the normalized conditional density matrix $\rho_{c|\ell_\text{max}}$ for $\ell_\text{max}=0$ and $\ell_\text{max}=1$ for the singlet (solid) and octet (dashed). Panel (b) shows the populations of the unnormalized blocks $\rho^{(\ell)}_c$, with the colors denoting $\ell=0,1$. The dotted horizontal lines in \textit{(a)} indicate the position space coherence $C_r$ of the $1\mathrm{S}$, $2\mathrm{S}$, and $3\mathrm{S}$ states.}
    \label{fig:3purification}
\end{figure*}

\textbf{\textit{Position-space coherence.---}}
Wigner negativity alone does not fully characterize the quantum structure of the density matrix, as states with a nonnegative Wigner function may still retain off-diagonal coherence. We therefore consider the relative entropy of coherence~\cite{Baumgratz:2013ecx,Streltsov:2016iow} in the position basis,
\begin{equation}
    C_r(\rho) = S(\text{diag}_r\rho) - S(\rho)
\end{equation}
as a measure of coherence, where $\text{diag}_r\rho$ denotes the density matrix obtained by setting all off-diagonal radial elements to zero in a discretized position basis. This provides a complementary measure of classicalization at the level of the density matrix. 
Note that the absolute value of $C_r$ depends on the position-space lattice spacing. The lattice-spacing-dependent contribution cancels in differences of $C_r$, so that the shape of its time evolution remains meaningful.
We show the time evolution of the coherence for the conditional density matrix $\rho_{c|\ell_\text{max}}$ in the singlet and octet sectors in panel (a) of Fig.~\ref{fig:3purification}. We observe that the coherence behaves similarly to the negativity, with a steep increase at the beginning of the evolution, strong damping by the medium, and a plateau at a finite value. The common origin of both behaviors can be traced to the structure of the dissipator. The jump operators $L_n$ of the pNRQCD master equation contain a term proportional to the position operator $r$ and a recoil correction $\mathrm{i}p/(2MT)$, thus showing a very similar structure to the Caldeira-Leggett master equation for quantum Brownian motion~\cite{CALDEIRA1983587}. The $r$-dependent contributions lead to a term proportional to $|r-r^\prime|^2\rho(r,r^\prime)$, which suppresses coherences between spatially separated configurations~\cite{Delorme:2024rdo,Brambilla:2025sis}. Under a Wigner transformation, the same contribution becomes momentum diffusion $\partial^2_p W_{\rho_c}$, which smoothes the negative regions of the Wigner distribution, thus suppressing negativity. If the recoil term is omitted, this diffusion is unopposed, and the leading-order evolution approaches an identity-like steady state with vanishing position-space coherence and Wigner negativity~\cite{Brambilla:2025sis}. The $p$-dependent contribution generates drag and establishes a diffusion–drag balance, in close analogy with the Lindblad form of the Caldeira–Leggett model~\cite{theoryofoqs,CALDEIRA1983587,Delorme:2024rdo}. This balance produces the observed finite late-time plateaus of both spatial coherence and Wigner negativity. 

\textbf{\textit{Conditional purification.---}}
To understand quantitatively the residual negativity and coherence, and the process of classicalization, we investigate the structure of the surviving population $\rho_{c|\ell_\mathrm{max}}$. The dipole interaction induces transitions to higher angular-momentum states and into the octet sector, so that the population of a low-lying angular-momentum sector $\mathrm{Tr}\rho^{(\ell)}_{c}$ generally decreases, as shown in panel (b) of Fig.~\ref{fig:3purification}.
In panel (c) of Fig.~\ref{fig:3purification}, the purity $P=\mathrm{Tr} \rho^2_{c|\ell_\mathrm{max}}$ shows an initial mixing followed by strong purification in the singlet sector, with $\rho_{s|0}$ approaching a pure state, while the octet remains mixed.
This purification is driven by the selection of modes, in particular, the dipole interaction predominantly dissolves weakly bound and unbound modes with a large radius, as the in-medium width scales with $\braket{r^2}$~\cite{Brambilla:2016wgg,Brambilla:2022ynh}. The surviving modes are therefore strongly dominated by the tightly bound states, mainly the ground state $1\mathrm{S}$. We refer to this process as \textit{conditional purification}, as it applies to the ensemble surviving in the selected color and angular momentum sector. We can see the conditional purification directly by investigating the overlap of the conditional density matrix with the bound state projectors $\mathrm{Tr}\left[ \rho_{s|0}\Pi_{n\mathrm{S}}\right]$ in panel (d) of Fig.~\ref{fig:3purification}. The overlap with the $1\mathrm{S}$ ground state increases and becomes close to one, indicating the purification of the conditional state. As the ensemble is dominated by the $1\mathrm{S}$ contribution, we expect the negativity and coherence to plateau close to the values of the $1\mathrm{S}$ bound state, which we observe in Fig.~\ref{fig:1_negativity} and panel (a) of Fig.~\ref{fig:3purification}.

\textbf{\textit{Conclusions.---}}
In this letter, we investigated the evolution and partial classicalization of bottomonium in the QGP. Within pNRQCD, we showed that the medium suppresses Wigner negativity and position-space coherence, which plateau at finite values close to those of the $1\mathrm{S}$ bound state.
The residual Wigner negativity precludes an exact interpretation of the phase-space distribution as a classical probability density and limits a direct reduction of the quantum master equation to Langevin dynamics. 

We further showed that the $2\mathrm{S}$ and $3\mathrm{S}$ overlaps receive sizable contributions from the negative regions in the Wigner transforms of the bound-state projectors, while the $1\mathrm{S}$ overlap is practically insensitive to these terms. Measured double ratios such as $R_{AA}[\Upsilon(3S)]/R_{AA}[\Upsilon(1S)]$ may therefore provide an observable handle on this quantum structure~\cite{CMS:2017ycw}. The semiclassical evolution itself was tested in the one-dimensional
abelian model of Ref.~\cite{Daddi-Hammou:2025hdz}, where it was found to describe the quantum yields well. That
calculation
follows the full relative density matrix and contains neither color nor
angular-momentum sectors, and therefore does not realize the conditional purification mechanism identified here. A direct comparison would require constructing the corresponding three-dimensional semiclassical evolution including the color dynamics of the pNRQCD master equation, which we leave to future work.

The observed partial classicalization arises from two distinct effects of the medium: Diffusion suppresses spatial coherence and Wigner negativity, while drag counteracts unrestricted diffusion and allows a nontrivial steady state. At the same time, the dipole interaction preferentially removes extended, weakly bound components, conditionally purifying the surviving singlet ensemble toward compact bound states dominated by the $1\mathrm{S}$. As the $1\mathrm{S}$ state itself retains coherence and Wigner negativity, this conditional purification naturally explains the finite late-time plateaus of both quantities.

\section*{Acknowledgments}
N.B. and T.M. thank Miguel Angel Escobedo for useful discussion. 
N.B. and T.M. acknowledge support by the DFG cluster of excellence ORIGINS funded by the Deutsche Forschungsgemeinschaft (DFG) under Germany's Excellence Strategy - EXC-2094-390783311. The work of N.B.
is supported by the DFG Grant No. BR 4058/5-1 ”Open quantum systems and effective
field theories for hard probes of hot and/or dense medium”. N.B. acknowledges the European Research Council advanced grant ERC-2023-ADG-Project EFT-XYZ.

\bibliographystyle{apsrev4-1}
\renewcommand*{\bibfont}{\footnotesize}
\bibliography{lit.bib}

\appendix
\clearpage
\onecolumngrid

\setcounter{equation}{0}
\setcounter{figure}{0}
\setcounter{table}{0}
\setcounter{page}{1}

\renewcommand{\theequation}{S\arabic{equation}}
\renewcommand{\thefigure}{S\arabic{figure}}
\renewcommand{\thetable}{S\arabic{table}}

\begin{center}
{\Large \textbf{Supplemental Material for ``How Quantum Is Bottomonium in the Quark-Gluon Plasma?''}}\\
Nora Brambilla, Tom Magorsch
\end{center}

\twocolumngrid

\section{The pNRQCD Lindblad equation}
\label{app:A}

We here provide the full form of the pNRQCD Lindblad equation given in Eq.~\eqref{eq:lind} for the bottomonium evolution. We consider the evolution at next-to-leading order in an expansion in the binding energy $E$ over the thermal scale $\pi T$, which was derived in Ref.~\cite{Brambilla:2022ynh}. In the spherical basis, the Hamiltonian for an angular momentum block $\ell$ takes the form 
\begin{align}
    H^\ell = \begin{pmatrix}
h^\ell_s + K_T & 0 \\
0 & h^\ell_o + \frac{N_{c}^{2}-2}{2\bigl(N_{c}^{2}-1\bigr)} K_T
\end{pmatrix},
\end{align}
with 
\begin{equation}
    K_T = \frac{r^{2}}{2}\gamma
+ \frac{\kappa}{4 M T}\{ r, p_{r} \}.
\end{equation}
The $2\times 2$ structure of the Hamiltonian reflects the color structure of the density matrix, corresponding to the singlet and octet color blocks.
Here $r$ is the radial position space operator, while $p_r$ denotes its conjugate momentum operator, $M$ is the bottom quark mass, $T$ is the temperature, and $N_c=3$ in QCD. The transport coefficients $\kappa$  and $\gamma$ parametrize the effects of the medium. The vacuum hamiltonians $h^\ell_s$ and $h^\ell_o$ are given by
\begin{equation}
    h^\ell_c = -\frac{1}{M}\left(\frac{\partial^2}{\partial r^2} + \frac{2}{r}\frac{\partial}{\partial r} \right) +V_c(r) + \frac{\ell(\ell+1)}{Mr^2},
\end{equation}
where $V_c(r)$ denotes the color singlet and color octet static potential. At leading order, these are given by an attractive and repulsive Coulomb potential, respectively, which we use in this study.

The jump operators $L_n$ induce transitions between color and angular momentum sectors. In total, there are six distinct jump operators corresponding to the processes of color transitions $s\to o$, $o\to s$, and $o\to o$, each with an angular momentum transition $\ell\to \ell+1$ and $\ell\to \ell-1$. 
In the spherical basis, the six jump operators are given by 
\begin{align}
    &L_{0} = \sqrt{\frac{\kappa}{N^2_c-1}}\begin{pmatrix}
        0 & 1 \\ 0 & 0 
    \end{pmatrix}O^{\downarrow}C^\downarrow_{o\to s},\\
    &L_{1} = \sqrt{\frac{\kappa}{N^2_c-1}}\begin{pmatrix}
        0 & 1 \\ 0 & 0 
    \end{pmatrix}O^{\uparrow}C^\uparrow_{o\to s},\\
    &L_{2} = \sqrt{\kappa}\begin{pmatrix}
        0 & 0 \\ 1 & 0 
    \end{pmatrix}O^{\downarrow}C^\downarrow_{s\to o},\\
    &L_{3} = \sqrt{\kappa}\begin{pmatrix}
        0 & 0 \\ 1 & 0 
    \end{pmatrix}O^{\uparrow}C^\uparrow_{s\to o},\\
    &L_{4} = \sqrt{\frac{\kappa(N^2_c-4)}{2(N^2_c-1)}}\begin{pmatrix}
        0 & 0 \\ 0 & 1 
    \end{pmatrix}O^{\downarrow} C^\downarrow_{o\to o},\\
    &L_{5} = \sqrt{\frac{\kappa(N^2_c-4)}{2(N^2_c-1)}}\begin{pmatrix}
        0 & 0 \\ 0 & 1 
    \end{pmatrix}O^{\uparrow}C^\uparrow_{o\to o},
\end{align}
with operators in position space
\begin{align}
    &C^\downarrow_{u\to v}=r\left(1+\frac{\Delta V_{uv}}{4T}\right)+\frac{1}{2MT}\left(\frac{\partial}{\partial r}+\frac{\ell+1}{r}\right),\\
    &C^\uparrow_{u\to v}=r\left(1+\frac{\Delta V_{uv}}{4T}\right)+\frac{1}{2MT}\left(\frac{\partial}{\partial r}-\frac{\ell}{r}\right),
\end{align}
where $\Delta V_{uv}= V_u(r) - V_v(r)$. The operators $O^{\uparrow\downarrow}$ induce transitions between angular momentum blocks and are given by
\begin{equation}
    O^\downarrow_{\ell^\prime,\ell} = \sqrt{\frac{\ell}{2\ell+1}} \delta_{\ell^\prime,\ell-1},\quad O^\uparrow_{\ell^\prime,\ell} = \sqrt{\frac{\ell+1}{2\ell+1}} \delta_{\ell^\prime,\ell+1}.
\end{equation}

In practice, we perform the simulations with the QTraj code~\cite{Omar:2021kra} with the setup described in Ref.~\cite{Brambilla:2022ynh}. We discretize position space in a box of $R_\text{max}=40\,\text{GeV}^{-1}$ on $N=2048$ lattice points. In the simulations, we implement a maximum angular momentum cutoff $\ell_\text{cut}$. If a trajectory reaches this $\ell_\text{cut}$, we assume that its probability to return to the low-lying states is sufficiently small that we can neglect it. We thus terminate these trajectories and set their contribution to the low-lying angular-momentum ensembles to zero. In this study, we choose $\ell_\text{cut}=12$. We checked that the survival probabilities have fully converged for this choice of $\ell_\text{cut}$. For the time evolution of the Lindblad equation, we use the Crank-Nicolson method implemented in QTraj with a time step of $\Delta t=\num{5e-6}\,\mathrm{GeV}^{-1}$, smaller than in previous studies. This finer step is required to accurately resolve the position-space structure of higher-angular-momentum states. For the low-lying angular-momentum sectors studied here, this is a conservative choice that does not affect the results.

\section{Wigner transform computation}
\label{app:B}

To compute the Wigner transformation~\eqref{eq:wig}, we decompose the density matrix in terms of spherical components. Since we assume isotropy, we can write the density matrix $\rho_c(\vec{r},\vec{r}^\prime)$ as 
\begin{equation}
    \rho_c(\vec{r},\vec{r}^\prime) = \frac{1}{4\pi} \sum_\ell \rho^{(\ell)}_c(r,r^\prime) P_\ell(\hat{r}\cdot \hat{r}^\prime),
\end{equation}
where $P_\ell$ are the Legendre polynomials and $\rho^{(\ell)}_c$ denotes the radial density summed over the $2\ell +1$ magnetic states so that $\mathrm{Tr}\rho_c = \sum_\ell \int r^2 \mathrm{d}r \rho^{(\ell)}_c(r,r)$. Furthermore for any vector $\vec{a}$ we use the notation $a=|\vec{a}|$ and $\hat{a}=\vec{a}/a$. The Wigner transformation thus takes the form 
\begin{align}
\begin{split}
    W_\rho(\vec{R},\vec{P}) = \frac{1}{2(2\pi)^4} \int &s^2 \mathrm{d}s\,\sin(\theta_s) \mathrm{d}\theta_s\,  \mathrm{d}\phi_s e^{-\mathrm{i}Ps\cos\theta_{Ps}}\\&\sum_\ell P_\ell(\hat{R}_-\cdot \hat{R}_+)\rho^{(\ell)}_c(R_-,R_+),
\end{split}
\label{eq:wigner_integral}
\end{align}
where $\vec{R}_\pm = \vec{R}\pm \vec{s}/2$, the coordinates $(s,\theta_s,\phi_s)$ denote the spherical coordinates of $s$ and $\theta_{Ps}=\angle(\vec{P},\vec{s})$. We can choose the coordinate system for the $\mathrm{d}s$ integral to have its $z$-axis along $\vec{R}$. In this case, we obtain 
\begin{align}
    R_\pm &= \sqrt{R^2 + s^2/4 \pm Rs\cos\theta_{s}}\\
    \hat{R}_-\cdot\hat{R}_+ &= \frac{R^2-s^2/4}{\sqrt{(R^2+s^2/4)^2 - R^2s^2\cos^2\theta_s}}.
\end{align}
The angle $\theta_{Ps}$ can be written as 
\begin{equation}
    \cos\theta_{Ps} = \sin\theta_{RP}\sin\theta_s\cos\phi_s +\cos\theta_{RP}\cos\theta_s.
\end{equation}
We can perform the $\phi_s$ integral in Eq.~\eqref{eq:wigner_integral} analytically which gives the full form 
\begin{align}
\begin{split}
    W_\rho(\vec{R},\vec{P}) = \frac{1}{2(2\pi)^3}\int& s^2\mathrm{d}s\sin\theta_s\mathrm{d}\theta_s e^{-\mathrm{i}Ps\cos\theta_{RP}\cos\theta_s}\\&J_0(Ps\sin\theta_{RP}\sin\theta_s)\\&\sum_\ell P_\ell(\hat{R}_-\cdot\hat{R}_+)\rho^{(\ell)}_c(R_-,R_+),
\end{split}
\end{align}
where $J_0$ is the zeroth-order Bessel function.
This expression contains two integrals over $\mathrm{d}s$ and $\mathrm{d}\theta_s$, which we perform numerically by discretizing the integration intervals and performing a simple trapezoid quadrature. We perform the integrals in this way using $\theta_s\in[0,\pi]$ with $N_{\theta_s}=1300$ points and $s\in[0,R_\text{max}]$ with $N_s=1000$. Here we use $R_\text{max}$ as the upper limit for the $s$-integral, although in principle $s$ is constrained by requiring both $\vec{R} \pm \vec{s}/2$ to lie within the box; since most the wave functions considered in this study decay well within $R_\text{max}$, the density matrices $\rho$ vanish before the boundary and in this case this choice captures the full support. This is only marginally satisfied for the projector of the $3\mathrm{S}$ state $\Pi_{3\mathrm{S}}$ and can lead to finite volume effects, on which we will comment below. 
We checked that the two integrals are converged in the discretizations $N_{\theta_s}$ and $N_s$. The resulting Wigner distribution is calculated on a grid of $N_R=500$, $N_P=400$, and $N_{\theta_{RP}}=512$ points, with $R\in [0,R_\text{max}],\theta_{RP}\in[0,\pi]$, and momentum spacing $\Delta P = \pi/R_\mathrm{max}$. Integrals over the Wigner distribution are calculated by using a trapezoidal quadrature in $(R,P,\theta_{RP})$.

\begin{figure}[t]
    \centering
    \includegraphics[width=0.9\linewidth]{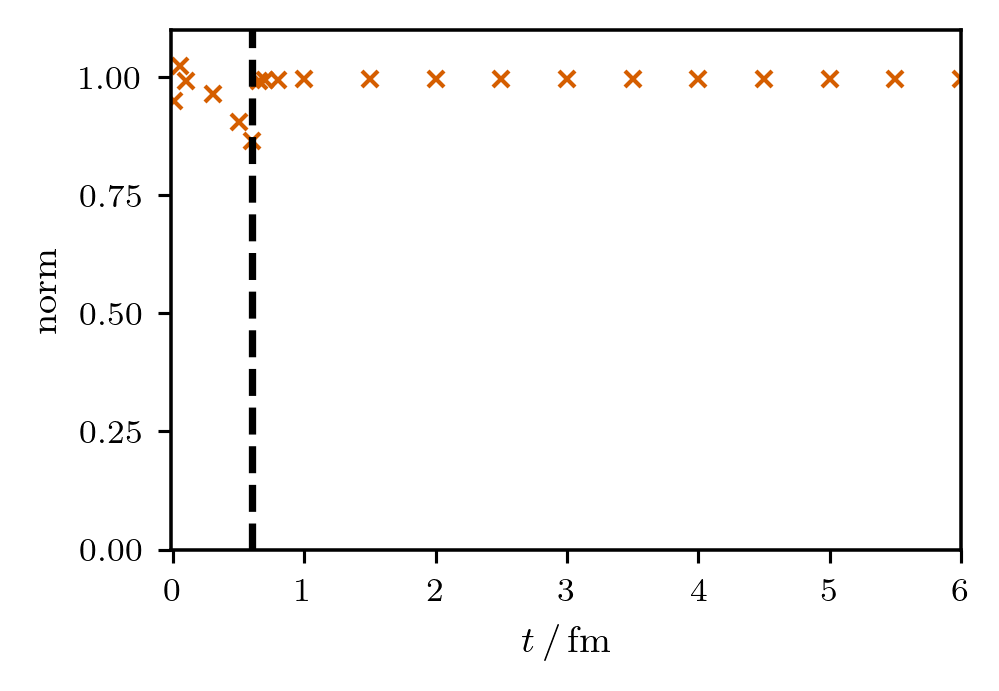}
    \caption{The time evolution of the norm of the Wigner distribution Eq.~\eqref{eq:Wigcond1}, calculated from the conditional density matrix $\rho_{s|0}$. As the conditional density matrix has trace one, the norm should be one, and any deviation stems from numerical inaccuracy.}
    \label{fig:normevolution}
\end{figure}
\begin{figure}[!t]
    \centering
    \includegraphics[width=0.99\linewidth]{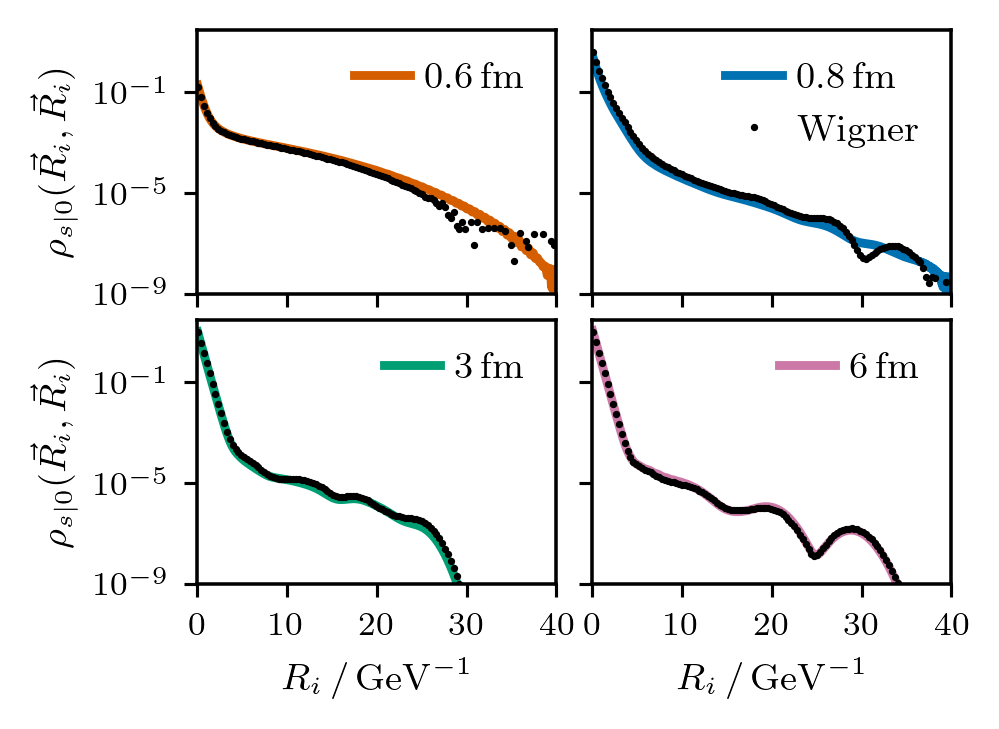}
    \caption{The diagonal of the conditional density matrix $\rho_{s|0}(\vec{R},\vec{R})$ at different times. We compare the exact diagonal of the density matrix (colored lines) to the result obtained by integrating over the Wigner distribution in Eq.~\eqref{eq:Wigcond2} (black dots). We plot only every third point from the Wigner distribution to keep the plots readable.}
    \label{fig:diagonal}
\end{figure}
As consistency checks we further tested the integration code performing on a Gaussian state $\rho^{(0)}(r,r^\prime) = 4\pi (\pi \sigma^2)^{-3/2}\exp\left[-\frac{1}{2\sigma^2}(r^2+{r^\prime}^2)\right]$, for which the Wigner distribution can be found analytically to be $W(R,P) = \frac{1}{\pi^3}e^{-R^2/\sigma^2 - P^2\sigma^2}$.
As a further consistency check for the numerical integration, we check the normalization conditions for the resulting Wigner distributions, which have to hold as
\begin{align}
   \int \mathrm{d}^3R \;\mathrm{d}^3P \,W(\vec{R},\vec{P}) &= \text{Tr}\rho \label{eq:Wigcond1}\\
   \int\mathrm{d}^3P \,W(\vec{R},\vec{P}) &= \rho(\vec{R},\vec{R}).\label{eq:Wigcond2}
\end{align}
As we perform the Wigner transformations of the conditional density matrix $\rho_{s|0}$ and projectors, we know that the trace should always be one. We thus use Eq.~\eqref{eq:Wigcond1} and Eq.~\eqref{eq:Wigcond2} as crosschecks for the numerical accuracy of the Wigner transformation. In Fig.~\ref{fig:normevolution} we show the norm of the Wigner distribution calculated by integrating over the data obtained from the conditional density matrix $\rho_{s|0}$. We observe that the norm is one with sub-percent accuracy during the Lindbladian evolution at $t>0.6\,$fm. In the vacuum evolution regime at early times $t<0.6\,$fm, the accuracy is slightly worse, with the largest deviations occurring at $0.6\,$fm right before the medium is turned on. We attribute this to finite volume effects, which enter through the finite range of the $\mathrm{d}s$-integration: since at $0.6\,$fm the wave function substantially evolved in position space and is thus larger at $R>20\,\mathrm{GeV}^{-1}$ than at other times and therefore carries off-diagonal support at separations $s$ approaching $R_\text{max}$.
\begin{figure}[!t]
    \centering
    \includegraphics[width=0.9\linewidth]{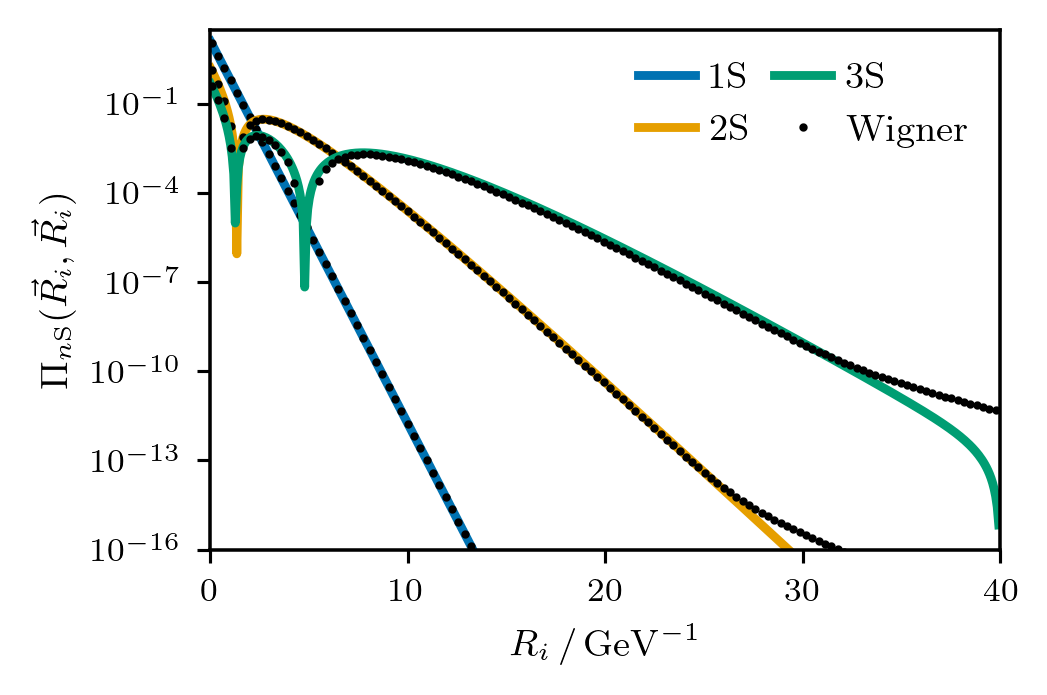}
    \caption{The diagonal of the bound state projectors $\Pi_{n\mathrm{S}}(\vec{R},\vec{R})$. We compare the exact diagonal of the projector (colored lines) to the result obtained by integrating over the Wigner distribution in Eq.~\eqref{eq:Wigcond2} (black dots). We plot only every third point from the Wigner distribution to keep the plots legible.}
    \label{fig:diagonal_bound}
\end{figure}
In particular increasing the integration range of the $\mathrm{d}s$ integral from $s_\text{max}=40\,\mathrm{GeV}^{-1}$ to $60\,\mathrm{GeV}^{-1}$, increases the norm of the Wigner distribution at $0.6\,$fm from $85\%$ to $91\%$. The negativity $\eta$, on the other hand, is less sensitive to these effects, as it is defined as a ratio of phase-space integrals, and only changes by about $1\%$. We likewise check Eq.~\eqref{eq:Wigcond2}, by numerically calculating the momentum integral over the Wigner distribution to obtain the diagonal of the conditional density matrix $\rho_{s|0}$, which we compare to the original data. We show the comparison for four different times in Fig.~\ref{fig:diagonal}.
We observe that at later times, there is perfect agreement across nine orders of magnitude. At $0.6\,$fm, however, there are deviations at $R>20\,\mathrm{GeV}^{-1}$, which we attribute to finite volume effects mentioned above. For $R<20\,\mathrm{GeV}^{-1}$, however, where the density matrix is largest, the Wigner distribution calculation agrees well with the original density matrix.
We performed the same checks for the bound state projectors $\Pi_{n\mathrm{S}}$. For the $1\mathrm{S}$ and $2\mathrm{S}$ projectors, we find the Wigner distribution norm to deviate from one by less than $1\%$. For the $3\mathrm{S}$ projector, however, we find a $13\%$ deviation, which we again attribute to finite volume effects, as the $3\mathrm{S}$ state has by far the largest radius of the three bound states. The conclusion for the 3S rests on the observation that $\mathrm{Tr}[\rho_{s|0}\Pi_3\mathrm{S}]$ is a small residue of the cancellation between the two large contributions $\mathcal{O}^{++}_{3\mathrm{S}}$ and $\mathcal{O}^{+-}_{3\mathrm{S}}$. This qualitative statement is unaffected by the limited numerical precision of the residue itself. We show the comparison of the diagonal of the density matrix to the Wigner integral for the three bound state projectors in Fig.~\ref{fig:diagonal_bound}.
We observe deviations from the exact diagonal for the $2\mathrm{S}$ and $3\mathrm{S}$ projectors at large $R$. Since the $3\mathrm{S}$ wave function is substantially larger at these large $R$ than the other bound states, finite volume effects are strongest for this state.

\section{Wigner distribution time evolution}
\label{app:C}

\begin{figure*}[t!]
    \centering
    \includegraphics[width=0.4\linewidth]
    {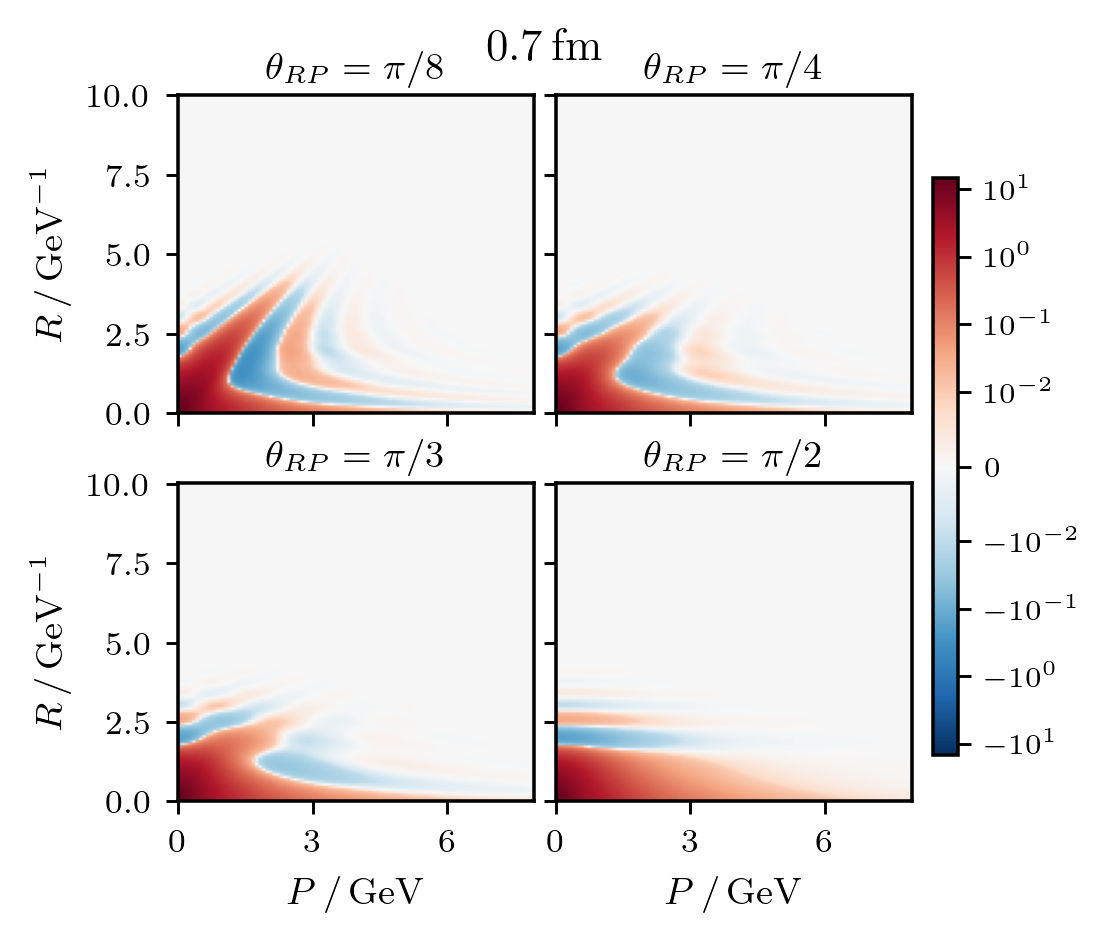}
    \includegraphics[width=0.4\linewidth]{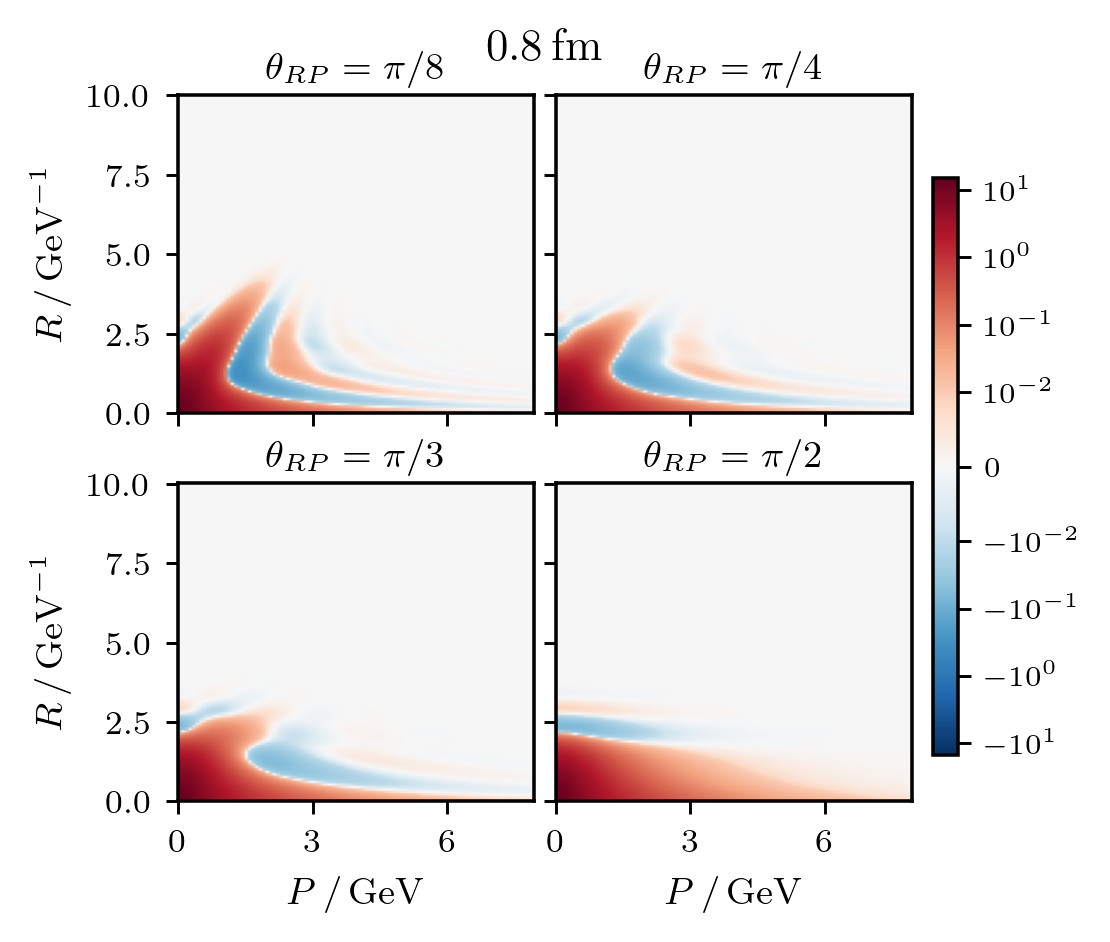}\\
    \includegraphics[width=0.4\linewidth]
    {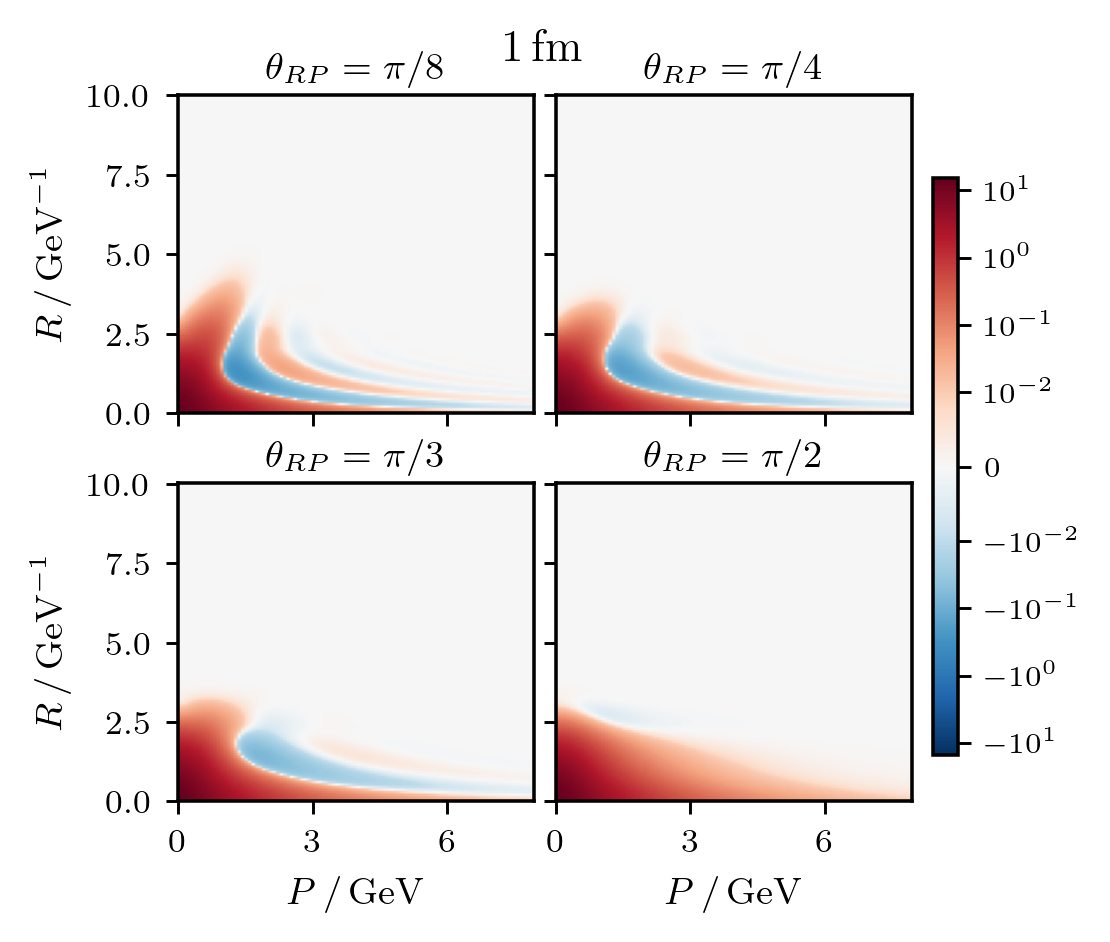}
    \includegraphics[width=0.4\linewidth]{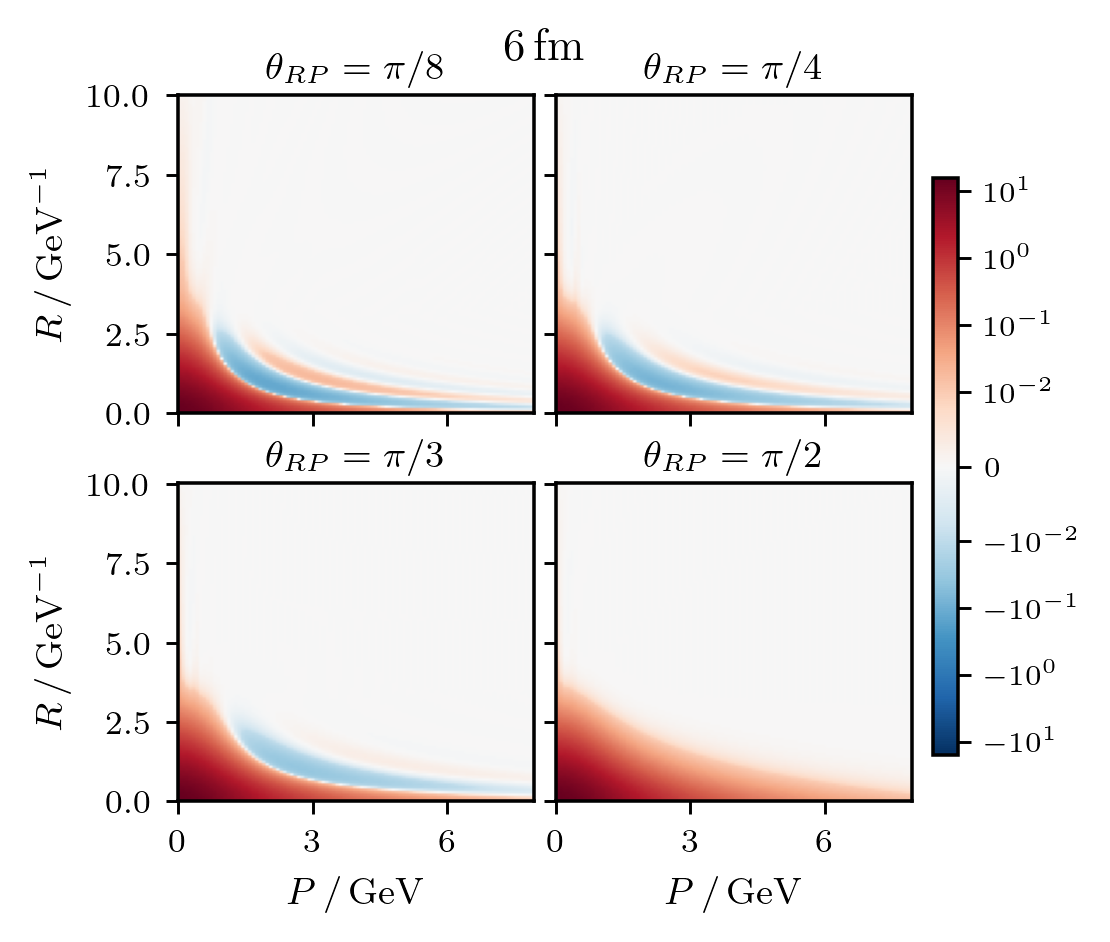}
    \caption{The Wigner transform $W(R,P,\theta_{RP})$ of $\rho_{s|0}(t)$ at four different angles $\theta_{RP}$ for different times $t$. From top left to bottom right: $t=0.7\,$fm, $0.8\,$fm, $1\,$fm and $6\,$fm.}
    \label{fig:S4_wig_0_7fm}
\end{figure*}

Here we show the explicit form of the Wigner transform $W(R,P,\theta_{RP})$, at different times. We plot the distribution for four different angles $\theta_{RP}$ at $t=0.7\,\mathrm{fm}$, $t=0.8\,\mathrm{fm}$, $t=1\,\mathrm{fm}$ and $t=6\,\mathrm{fm}$ in Fig.~\ref{fig:S4_wig_0_7fm}.

\begin{figure*}[t]
    \centering
\includegraphics[width=0.4\linewidth]{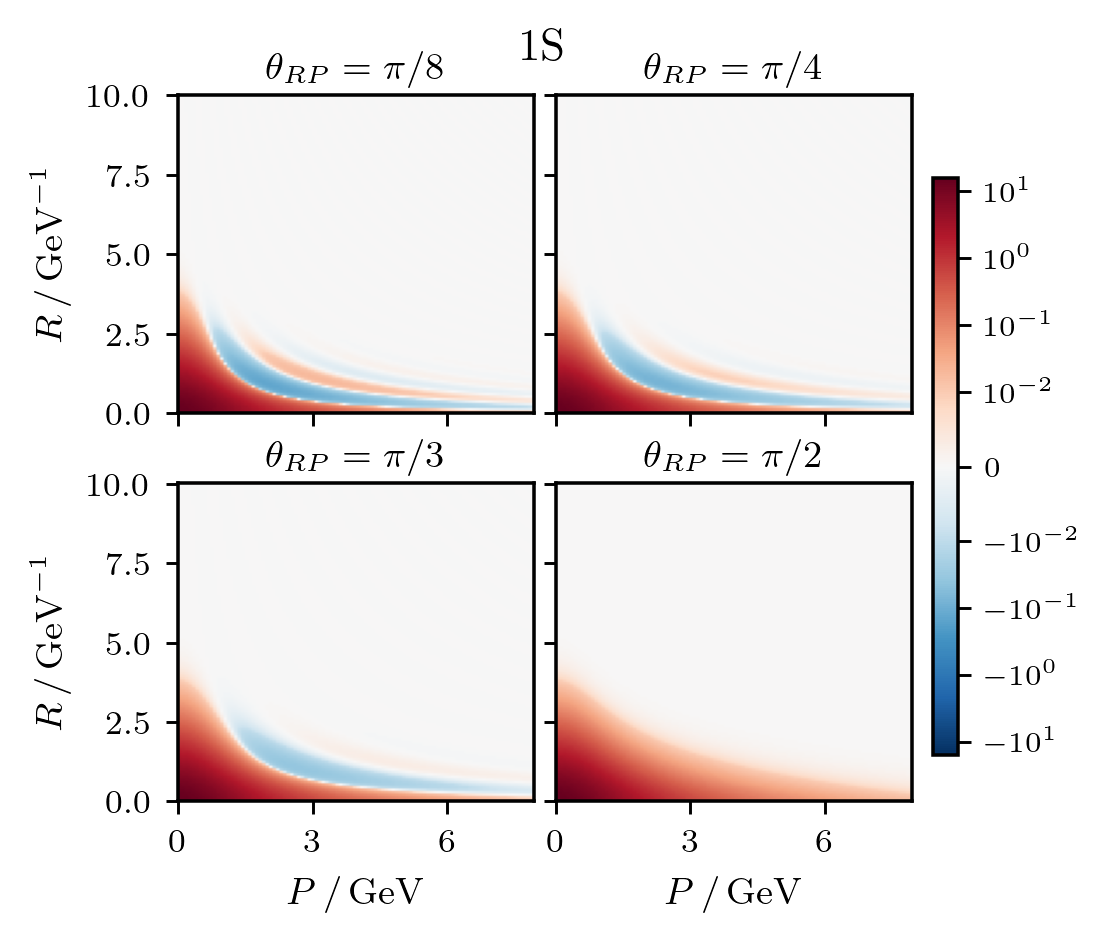}
\includegraphics[width=0.4\linewidth]{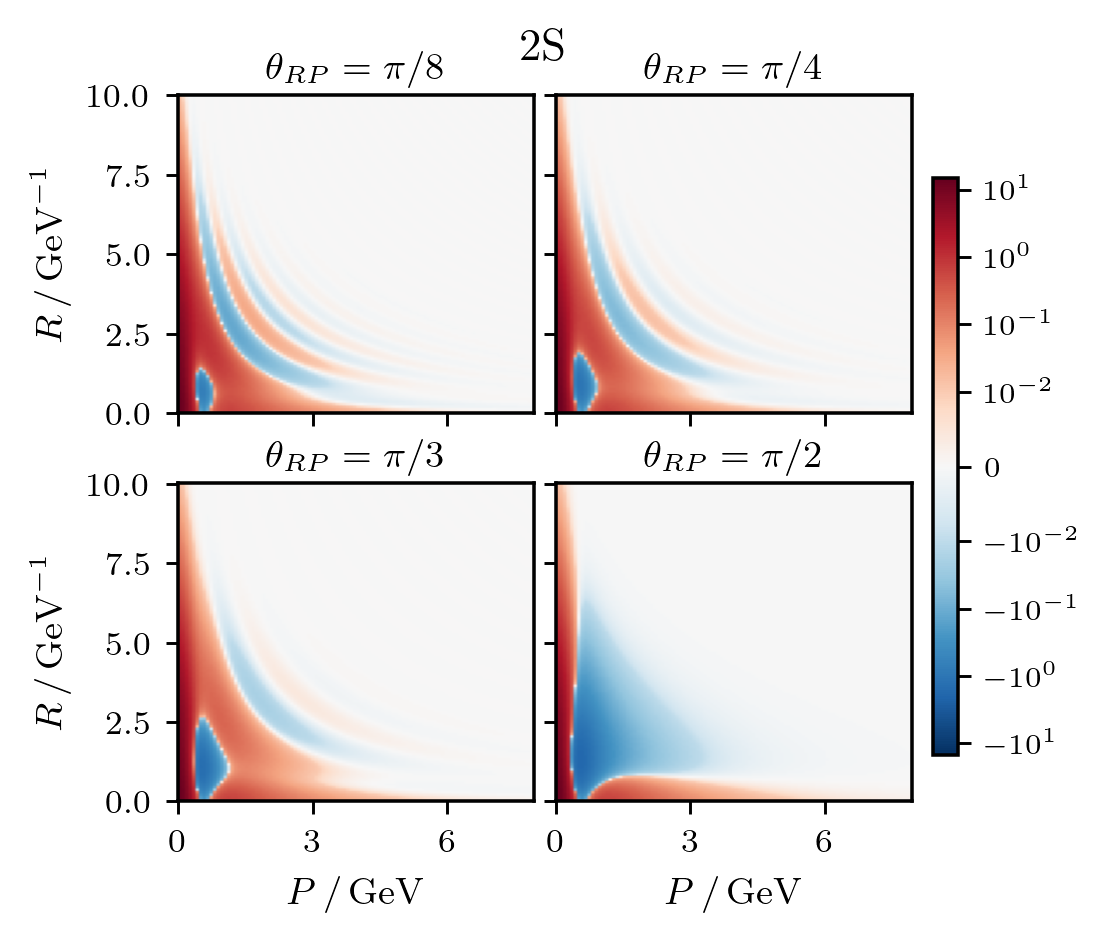}\\
\includegraphics[width=0.4\linewidth]{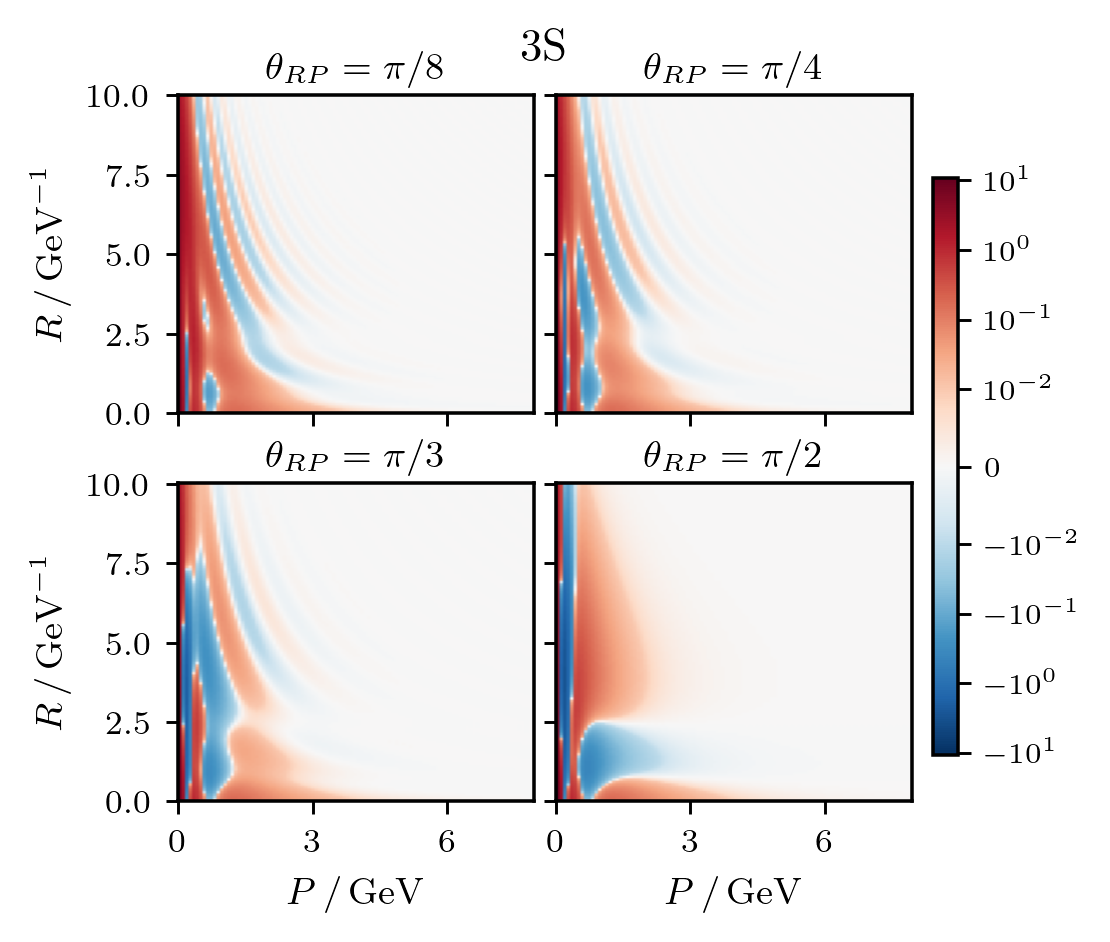}
    \caption{The Wigner transform $W(R,P,\theta_{RP})$ for the $1\mathrm{S}$ \textit{(top left)}, $2\mathrm{S}$ \textit{(top right)} and $3\mathrm{S}$ \textit{(bottom)} projectors $\Pi_{n\mathrm{S}}$ at four different angles $\theta_{RP}$.}
    \label{fig:S4_1S}
\end{figure*}

At all times, we observe negative regions in the Wigner distribution, leading to the observed negativity reported in the main text. Comparing the Wigner distribution at early times with that at $6\,$fm, we see that the negative regions have been smoothed out, as expected from the Lindbladian evolution of the diffusion term. We can compare the Wigner distribution of the time-evolved density matrix with those from the bound-state projectors $\Pi_{n\mathrm{S}}$, as shown in Fig.~\ref{fig:S4_1S}.

We notice that all bound states show visible negative fringes. For the $2\mathrm{S}$ and $3\mathrm{S}$ there are larger negative regions, especially for large $\theta_{RP}$. Furthermore, we can observe that the $1\mathrm{S}$ Wigner distribution is very similar to the Wigner distribution of the density matrix at $6\,$fm in Fig.~\ref{fig:S4_wig_0_7fm}. This reflects the conditional purification of $\rho_{s|0}$, which evolves very close to a $1\mathrm{S}$ state.

\section{Entropy evolution}
\label{app:D}
\begin{figure}[t]
    \centering
    \includegraphics[width=0.9\linewidth]{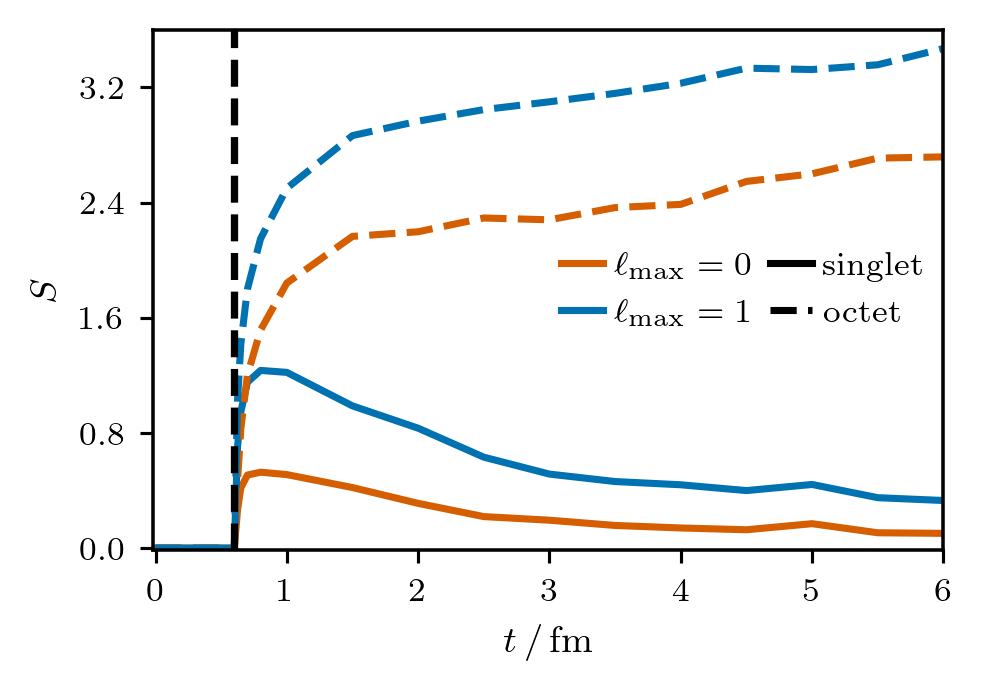}
    \caption{The time evolution of the entropy $S$ for the conditional density matrix $\rho_{c|\ell_\text{max}}$. The singlet lines show the entropy for the conditional density matrix in the singlet sector, while the dashed lines depict the entropy for the octet sector.}
    \label{fig:entropy}
\end{figure}

In Fig.~\ref{fig:entropy} we show the entropy $S[\rho_{c|\ell_\text{max}}]=-\mathrm{Tr}[\rho_{c|\ell_\text{max}}\ln\rho_{c|\ell_\text{max}}]$ for the conditional density matrix $\rho_{c|\ell_\text{max}}$. Naturally, the entropy is expected to behave in a conjugate
way with respect to the purity studied in the main text. The entropy is zero during the initial vacuum evolution, as the state is pure; it then rises significantly when the medium is turned on due to strong initial mixing in both the singlet and octet sectors. In the singlet sector, it then decreases after a maximum at $1\,$fm, while the octet keeps increasing. The fact that the entropy slowly reaches a
plateau is consistent with the onset of equilibration of the ensemble.
We stress that the decrease does not indicate that the bottomonium as a whole becomes less mixed. It reflects the conditional purification: the dipole interaction transfers the bulk of the population to higher angular momentum, and the conditional density matrix retains only the small surviving low-$\ell$ fraction, which lies close to the pure $1\mathrm{S}$ state. We emphasize that this does not make a statement about the entropy of the full density matrix $\rho$, which is not guaranteed to increase monotonically under the Lindblad evolution~\eqref{eq:lind}.

\end{document}